\documentclass[sigconf]{acmart}

\usepackage{pdfpages}
\pdfoutput=1
\usepackage{multirow}
\usepackage{booktabs}

\def\BibTeX{{\rm B\kern-.05em{\sc i\kern-.025em b}\kern-.08emT\kern-.1667em\lower.7ex\hbox{E}\kern-.125emX}}
    
\acmConference[]{}{}{} 
\acmYear{2019}
\copyrightyear{2019}

\begin{document}

%
% The "title" command has an optional parameter, allowing the author to define a "short title" to be used in page headers.
\title{Document Network Embedding: Coping for Missing Content and Missing Links}

%
% The "author" command and its associated commands are used to define the authors and their affiliations.
% Of note is the shared affiliation of the first two authors, and the "authornote" and "authornotemark" commands
% used to denote shared contribution to the research.

\author{Jean Dupuy}
\affiliation{%
  \institution{Universit\'e de Lyon, Lyon 2, ERIC EA3083 -- MeetSYS}
  %\city{Lyon}
  %\state{France}
}
\email{jean.dupuy@meetsys.com}

\author{Adrien Guille}
\affiliation{%
  \institution{Universit\'e de Lyon, Lyon 2, ERIC EA3083}
  %\institution{University of Lyon}
  %\city{Lyon}
  %\state{France}
}
\email{adrien.guille@univ-lyon2.fr}

\author{Julien Jacques}
\affiliation{%
  \institution{Universit\'e de Lyon, Lyon 2, ERIC EA3083 }
  %\institution{University of Lyon}
  %\city{Lyon}
  %\state{France}
}
\email{julien.jacques@univ-lyon2.fr}

%
% By default, the full list of authors will be used in the page headers. Often, this list is too long, and will overlap
% other information printed in the page headers. This command allows the author to define a more concise list
% of authors' names for this purpose.
\renewcommand{\shortauthors}{J. Dupuy, A. Guille and J. Jacques}

%
% The abstract is a short summary of the work to be presented in the article.
\begin{abstract}
Searching through networks of documents is an important task. A promising path to improve the performance of information retrieval systems in this context is to leverage dense node and content representations learned with embedding techniques. However, these techniques cannot learn representations for documents that are either isolated or whose content is missing.
To tackle this issue, assuming that the topology of the network and the content of the documents correlate, we propose to estimate the missing node representations from the available content representations, and conversely. Inspired by recent advances in machine translation, we detail in this paper how to learn a linear transformation from a set of aligned content and node representations. The projection matrix is efficiently calculated in terms of the singular value decomposition.  
%
%This approach proves relevant, as evidenced by the quality of the alignment of the node representations with the projected content representations. 
The usefulness of the proposed method is highlighted by the improved ability to predict the neighborhood of nodes whose links are unobserved based on the projected content representations, and to retrieve similar documents when content is missing, based on the projected node representations.
%That shows it is indeed possible to gain structural knowledge about a node, solely from its content.
\end{abstract}



%
% Keywords. The author(s) should pick words that accurately describe the work being
% presented. Separate the keywords with commas.

%
% This command processes the author and affiliation and title information and builds
% the first part of the formatted document.
\maketitle

\section{Introduction}

Searching through a network of documents, such as scientific articles \cite{brochier2019phd}, Web pages or social media posts \cite{Guille:2013:IDO:2503792.2503797}, is a common task and a long-standing research topic. Recent works propose to improve the performance of information retrieval systems by leveraging dense vector space representations of the content and the nodes \cite{wen2018recommendation, seyler2018ri, seyler2018ri, zhang2019neural, makarov2018coauthorship}. Quite often, for various reasons, such as technical issues or legal limitations (\textit{e.g.} paywall), not all links can be observed nor can all the content be retrieved.

These representations are learned with algorithms similar to those devised for word embedding, \textit{e.g.} DeepWalk \cite{perozzi2014deepwalk}, Node2Vec \cite{grover2016node2vec} or GVNR \cite{brochier2019gvnr} for node representations, or Doc2Vec \cite{le2014paragraph} for content. Obviously, missing content prevents applying any document embedding techniques. Also, since network embedding algorithms perform random walks to measure the co-occurrence frequency between nodes, they cannot learn representations of isolated nodes.

In this paper, we address this issue by formulating a translation task, from available content representations to node representations, and from available node representations to content representations. Assuming that the topology of a network of documents correlates with the content of the documents, we propose to estimate the missing node (resp. document) representations as a linear transformation, \textit{i.e.} projection, of the content (resp. node) representations. More specifically, we show how to construct a dictionary of pairs of content and node representations that allows learning a projection matrix, inspired by recent advances in machine translation with pre-trained word embeddings \cite{mikolov2013translation, Xing2015NormalizedWE, smith2017orthogonal}. Note that a self-translation network embedding algorithm, STNE \cite{lie2018stne}, has been recently proposed, however it addresses a different task. It learns to translate a sequence of content to a sequence of nodes to learn richer node representations and it cannot operate on isolated nodes nor deal with missing content.

Our approach proves relevant, as evidenced by the quality of the alignment between the estimated representations and the true representations. Practically speaking, our method has two main applications. On the one hand, estimating the node representation helps in recovering the missing links. On the other hand, estimating the content representation from the node representation allows measuring the similarity with the content of other documents and thus provides helpful clues about the missing content.

%In particular, we think this method could help improving the performance of information retrieval systems that are based on network embedding techniques.

The rest of this paper is organized as follows. In section \ref{sec:related_work}, we survey related work on document and network embedding, as well as embedding-based bilingual translation. We then describe our proposal to estimate the missing representations in section \ref{sec:proposal}. Next, in section \ref{sec:xp}, we evaluate the quality of the projection and assess the improvement brought by our method in predicting the neighborhood of nodes whose links are missing and retrieve documents similar to those without content. Finally, we conclude and provide future directions in section \ref{sec:conclusion}.

\section{Related Work}
\label{sec:related_work}

Word embedding, \textit{i.e.}, the task of learning dense vector space representations of words, is tightly connected to the tasks of learning document and node representations, \textit{i.e.} document embedding and network embedding. In this section, we first briefly survey works related to word embedding and then present models adapted for document and network embedding. Lastly, we cover recent advances in offline bilingual translation via word embeddings, on which our work is based.

\subsection{Word Embedding}

Word embedding builds upon the distributional hypothesis. It states that distributional similarity and meaning similarity are correlated, which allows learning representations of words based on the contexts in which they occur, the context of a word being co-occurring words in a short window.

Mikolov \textit{et al.} \cite{mikolov2013efficient} propose two models, namely Skip-Gram and Continuous Bag-of-words (CBOW). The first models the conditional probability of occurrence of a word, given a word in the context. The second models the conditional probability of occurrence of a word, given the set of words in the context. In both cases, the probability is defined as a softmax function parameterized by the dot products of the vector representations of the words.

Because the softmax makes the estimation of the word representations computationally expensive, Mikolov \textit{et al.} \cite{mikolov2013distributed} propose approximate solutions based either on the hierarchical softmax or negative sampling, an approximation akin to noise-contrastive estimation.

\subsection{Content Representation Learning}

Le and Mikolov \cite{le2014paragraph} extends the Skip-Gram and CBOW models to learn representations of documents, under the names, respectively, Distributed Bag-of-words (DV-DBOW) and Distributed Memory (DV-DM). They rely on a simple trick, that consists in considering each document as a distinct new vocabulary entry, that occurs in every window in the document. Thus, these models learn document representations that are good at predicting the words they contain.

\subsection{Node Representation Learning}

Even though the distributional hypothesis originated in linguistics, Perozzi \textit{et al.} \cite{perozzi2014deepwalk} show that sequences of nodes generated by random walks and sentences share some statistical properties. In particular, they show that the frequency at which nodes appear in short random walks follows a power-law distribution, like the frequency of words in language. Consequently, they propose to learn node representations under the distributional hypothesis. The proposed model, DeepWalk, consists in learning the representations based on the Skip-Gram model with the hierarchical softmax approximation, from a corpus of node sequences -- deemed equivalent to sentences -- generated by truncated random walks.

LINE \cite{tang2015line} refines the objective optimized by DeepWalk, trying to preserve both the first and second order proximities in the embedding space. The representations are learned based on the Skip-Gram model with negative sampling rather than the hierarchical softmax approximation, using a specific stochastic gradient descent algorithm for weighted networks, that samples the edges according to their weights.

Node2vec \cite{grover2016node2vec} performs biased random walks, in order to better balance the exploration-exploitation trade-off, arguing that the added flexibility in exploring neighborhoods helps learning richer representations. The representations are also learned based on the Skip-Gram model with negative sampling.

\subsection{Bilingual Translation}

Mikolov \textit{et al.} \cite{mikolov2013translation} first propose to learn a linear transformation from the embedding space of a source language into the embedding space of a target language. They suggest to estimate a projection matrix from pre-trained vectors and a bilingual dictionary. To this end they formulate a least square problem based on the Euclidean distance between aligned vectors, which they solve with the regular stochastic gradient descent algorithm.

Even though this approach yields encouraging results, Xing \textit{et al.} \cite{Xing2015NormalizedWE} argue that the problem is ill-posed, since the similarity between representations is usually measured in terms of the cosine similarity, rather than the Euclidean distance. They show how to learn unitary word embeddings and formulate a maximization problem in terms of the dot product between some source vectors and projected target vectors. They also require the projection matrix to be orthogonal, so that the projected vectors remain unitary in order to preserve the equivalence between the dot product and the cosine similarity. Eventually, they devise a gradient descent based algorithm to learn the projection matrix. Yet, their approach is computationally expensive, because it requires to orthogonalize the projection matrix after each update, by computing its singular value decomposition.

Smith \textit{et al.} \cite{smith2017orthogonal} show that finding the projection matrix that solves the problem stated by Xing \textit{et al.} is actually equivalent to solving a least square problem based on the Euclidean distance, when the vectors are unitary and the projection matrix is orthogonal. Furthermore, they link this problem with the orthogonal Procrustes problem, which admits an analytical solution. The projection matrix is obtained from the singular value decomposition of the similarity matrix between the aligned source and target vectors. 

\section{Proposed Method}
\label{sec:proposal}

In this section, we describe our proposal, which consists in adapting techniques devised for bilingual translation to learn a projection from the content representations (pre-trained with, \textit{e.g.} Doc2Vec) to the node representations (pre-trained with, \textit{e.g.} DeepWalk), and vice versa.

Consider a corpus of documents organized in a network, so that the content of some documents is missing and the links of some other documents are unobserved. We assume a set of pre-trained node representations $A = \{a_i \in \mathbb{R}^k\}$ and a set of document representations $B = \{b_i \in \mathbb{R}^k\}$. Our goal is to estimate the missing node representations in $A$, from the content representations in $B$ ; conversely, estimate the missing content representations in $B$, from the node representations in $A$.

We posit that the topology of the network and the content of the documents correlate. Thus, we aim at finding a projection matrix $W \in \mathbb{R}^{k \times k}$ that maps content representations to node representations, and node representations to content representations. More particularly, given a document $i$, we want $W$ to maximize the cosine of the angle $\theta_{a_i, b_iW}$ between $a_i$ and $b_iW$. That is to say, we want to find a linear transformation so that the vectors $a_i$ and $b_iW$ point in the same direction. More formally, based on a set of pairs of node representations and document representations, $S$, we want to find a matrix $W$ so that:
\begin{equation}
\max_{W} \sum_{i \in S} \cos (\theta_{a_i, b_iW}),
\label{form:max1}
\end{equation}
In the rest of this section, we describe how to construct $S$ and then detail how to learn the projection matrix $W$.

\subsection{Dictionary construction}

Rather than learning the projection from all the available pairs of node representations and content representations, we suggest to restrain to a specific subset. The aim in doing so is (i) favor the learning of a good projection and (ii) limit the computational cost. 

Bojanowski \textit{et al.} \cite{bojanowski2017enriching} note that Skip-Gram, the model behind most network embedding algorithms, struggles to learn good representations for rare words. This also holds for network embedding algorithms, since Perrozi \textit{et al.} \cite{perozzi2014deepwalk} show that the frequency at which nodes occur in short random walks and the frequency of words in language follow a similar law. Hence, we propose to avoid rare nodes and to construct the dictionary only from the $m$ nodes that occur the most in short random walks on the network. Formally, with $f_i$ the frequency of node $i$, we construct the set $S$ of cardinality $m$ that maximizes $\sum_{i \in S} f_i$.

\subsection{Projection Learning}

Following \cite{Xing2015NormalizedWE}, we rewrite formula \ref{form:max1} in terms of the dot product, and thus require the node and content representation to be unitary vectors and constrain $W$ to be orthogonal:
\begin{equation}
\max_{W} \sum_{i \in S} a_i \cdot b_iW \text{, subject to}~ W^\top W = I.
\label{form:max2}
\end{equation}
Yet, rather than learning the representations again as Xing \textit{et al.} suggest, we simply row-normalize $A$ and $B$. Then, thanks to the proportional relationship between the opposite of the dot product and the squared Euclidean distance highlighted in \cite{smith2017orthogonal}, because $|| a_i - b_iW ||^2 = ||a_i||^2 + ||b_i||^2 - 2(a_i \cdot b_iW)$, we transform the problem into a minimization one:
\begin{equation}
\min_W \sum_{i \in S} || a_i - b_iW ||^2 \text{, subject to}~  W^\top W = I.
\end{equation}
This is a classic linear algebra problem, which solution is given in \cite{Schonemann1966procustres}. It consists in calculating the singular value decomposition of the matrix $M = A_S^\top {B_S}$, with $A_S \in \mathbb{R}^{m \times k}$ and $B_S \in \mathbb{R}^{m \times k}$ the aligned node and content representations according to the dictionary $S$. With $M = U \Sigma V^\top$, the matrix $W$ is defined as:
\begin{equation}
 W = UV^\top.
\end{equation}
We estimate the node representation of an isolated node by calculating $\tilde{a} = bW$. Similarly, we estimate the content representation of a document whose content is missing by calculating $\tilde{b}=a W^\top$.

%
%Note that, to reduce the computation time, one can actually limit $U$ and $V$ to the $d$ first left and right singular vectors and consider the rank-$d$ approximation $W_d = U_d {V_d}^\top$. This approximation can be efficiently computed in $\mathcal{O}\big(m^2 \log(d)\big)$ using the algorithm described in \cite{halko2011randomized}.

\section{Experiments}
\label{sec:xp}

In this section, we evaluate our proposal on two networks of documents. First, we focus on missing links and show that estimating the node representations from the content representations allows to better reconstruct the neighborhood of nodes whose links are unobserved.
Second, we focus on missing content and show that estimating the content representations from node representations allows to retrieve a significant fraction of the similar documents.

\subsection{Experimental setup}

\subsubsection{Data}

We consider two well known networks of scientific articles, Cora \cite{McCallum2000} and HepTh \cite{Leskovec2005}, whose properties are summed up in Table \ref{tab:networks}. No content nor citation links are missing. These datasets available online\footnote{Get the data: \url{https://github.com/thunlp/CANE/tree/master/datasets/cora}, \url{https://github.com/thunlp/CANE/tree/master/datasets/HepTh}}.

\begin{table}[]
\caption{Network properties.}
\begin{tabular}{r c c l}
\toprule
Name & \# of documents & \# of links  \\
\midrule
Cora & 2,211 & 5,001 \\
HepTh & 1,038 & 1,974\\
%New York Times & 5,135 & 3,050,513 \\
\bottomrule
\end{tabular}
\label{tab:networks}
\end{table}

\subsubsection{Node and Content Representations}

We experiment with DeepWalk \cite{perozzi2014deepwalk} to learn the node representations, with the following configuration: 80 random walks length 80 from each node, and a co-occurrence window of size 10. Regarding content representation, we experiment with the Distributed Memory (DV-DM) variant of Doc2Vec \cite{le2014paragraph} with a window of size 10. Here after, we report the results for representations of size 500.

\subsubsection{Projection Learning}

We learn the projection matrix via the exact singular value decomposition, based on a dictionary of size 1400 for Cora, and a dictionary of size 550 for HepTh.

\subsubsection{Similarity Metric}

We measure the similarity between vector representations in terms of cosine similarity \cite{mikolov2013efficient}.

\subsection{Missing Links: Neighborhood Reconstruction} 

We begin by assessing the quality of the estimated node representations. The evaluation task consists in reconstructing the neighborhood of a document whose links are unobserved. We adopt a leave-one-out strategy and proceed in the following manner. First, we learn all the content representations from the entire corpus. Then, for each document $i$, we do the following:

\begin{enumerate}
    \item Hide all the incoming and outgoing links of document $i$;
    \item Learn the node representations on this sub-network;
    \item Learn the projection matrix on this sub-network;
    \item Measure the similarity between the estimated node representation $\tilde{a}_i = b_iW$ and all the other node representations $\{a_j\}$;
    \item Order the documents accordingly and measure the precision at $n$, $P@n$ w.r.t. the true links.
\end{enumerate}
We report the relative gain in average precision in Table \ref{tab:link}, w.r.t. two baselines:

\begin{itemize}
    \item $\cos (\theta_{b_i, b_j})$: Documents ordered according to the similarity between the content representations;
    \item Random$_{200}$: Documents ordered randomly, picked among the 200 documents with the highest degrees.
\end{itemize}
It reveals that the estimated node representations, obtained via the linear transformation of the content representations always lead to a better precision in predicting the neighborhood of a document.

\begin{table}[]
\caption{Precision gain in link prediction.}
\begin{tabular}{cccccc}
\toprule
&  & P@5 & P@10 & P@20 & P@50 \\
\midrule
\multirow{2}{*}{Cora} &  $\cos (\theta_{b_i, b_j})$ & +4.4\% & +23.4\% & +20.1\% & +20.0\% \\
 & Random$_{200}$ & +681.4\% & +475.9\% & +166.1\% & +51.9\% \\
\midrule
\multirow{2}{*}{Hepth} & $\cos (\theta_{b_i, b_j})$ & +69.2\% & +42.9\% & +5.9\% & +2.6\% \\
 & Random$_{200}$ & +516.6\% & +375.1\% & +350.4\% & +175.6\% \\
\bottomrule

\end{tabular}
\label{tab:link}
\end{table}

\subsection{Missing Content: Similar Document Retrieval}

We now switch our attention to missing content. We assess the quality of the estimated content representations following a similar leave-one-out methodology. We learn all the the content representations and define the true sets of similar document $S_i = \{j | \cos (\theta_{b_i, b_j}) > 0.2\}$. The evaluation task consists in reconstructing the set of similar documents for a document whose content is unknown. We proceed as follows. First, we learn all the node representations from the whole network. Then, for each document $i$, we do the following:

\begin{enumerate}
    \item Hide document $i$ from the corpus;
    \item Retrain the content representations on this sub-corpus;
    \item Learn the projection matrix on this sub-corpus;
    \item Measure the similarity between the estimated content representation $\tilde{b}_i = a_iW^\top$ and all the content representations $\{b_j\}$;
    \item Measure the accuracy by comparing the true set of similar documents $S_i$ and the estimated set of similar documents $\tilde{S}_i = \{j | \cos (\theta_{\tilde{b}_i, b_j}) > 0.2\}$.
\end{enumerate}
We report the average precision following this procedure in the first row of Table \ref{tab:content}. The second row shows the average precision based on the node representations, where $\tilde{S}_i = \{j | \cos (\theta_{a_i, a_j}) > 0.2\}$. It reveals that estimating the missing content representation from the corresponding node representation doubles the accuracy in document retrieval, in comparison with the accuracy obtained by measuring the similarity based on the node representations. In particular, we are able to recover, on average, two-thirds of the similar documents in the Hepth corpus.

\begin{table}[]
\caption{Average accuracy in similar document retrieval.}
\begin{tabular}{r cc }
\toprule
 & Cora &  Hepth  \\
\midrule
$\tilde{S}_i = \{j | \cos (\theta_{\tilde{b}_i, b_j}) > 0.2\}$ & \textbf{0.36}  & \textbf{0.65}  \\
$\tilde{S}_i = \{j | \cos (\theta_{a_i, a_j}) > 0.2\}$ & 0.13 &  0.30  \\
\bottomrule
\end{tabular}
\label{tab:content}
\end{table}

%We also evaluate our proposal on the reversed task. We used a similar protocol to the link prediction task developed in the previous section, but we removed the content of a document instead of its node in the graph. 
%For all content removed, we look for the true $n$ closest documents in th content embedding. We present recalls between this set and the set of the $n$ closest documents to the projected node corresponding to the removed content in the content space ($r_{projected}$), and then with the set of the $n$ closest documents of the missing document in the node embedding space ($r_{node}$).

%In addition we extend this evaluation method to all documents by using cosine similarity with the missing content document greater than 0.3. Indeed, as all documents are not close to each others it could be hard to retrieve the most dissimilar ones. We find that projecting node representation of the missing content document help to retrieve $63.9\%$ of these documents on Cora, and $79.54\%$ on HepTh.

%Results in Table \ref{tab:recall} show the using the linear projection increase our capacity to retrieve similar documents to an unobserved content if we only know its node representation.

\section{Conclusion}
\label{sec:conclusion}

In this paper, we described a way to efficiently learn a linear function to map pre-trained node and content representations learned from a network of documents. In presence of missing links and missing content, this method helps in reconstructing the neighborhood of isolated documents and also helps in recovering documents that are similar to a document whose content is unknown. In future work, we would like to investigate how to learn a non-linear transformation to refine the mapping between the two spaces.
%Incorporating this simple method in an existing information retrieval system that relies on network embedding would help improve its performance, by allowing it to consider more candidate nodes. Another direct applications would be recommending links for a new document. It also suggests that it could approximate an unseen content representation from node representation, which would help content-based recommendation in incomplete document networks. 
%In future work, we plan to investigate the applications of the reciprocal projection, from node representations to content representations.

\bibliographystyle{ACM-Reference-Format}
\bibliography{references}

\end{document}